\documentclass{article}

\begin{document}
\title{Revisiting the calculation of inflationary perturbations.$^1$}
\author{C\'esar A. Terrero-Escalante$^{\dagger}$$^2$, 
Dominik J. Schwarz$^*$$^3$, \\
and \\
Alberto A. Garc\'{\i}a$^{\dagger}$$^4$ \\[12pt]
$\dagger$ Departamento~de~F\'{\i}sica,\\
~Centro de Investigaci\'on y de Estudios Avanzados del IPN,\\
~Apdo.~Postal~14-740,~07000,~M\'exico~D.F.,~M\'exico.\\
$*$ Institut f\"ur Theoretische Physik, TU-Wien,\\
Wiedner Hauptstra\ss e 8--10, A-1040 Wien, Austria.}
\maketitle
\footnotetext[1]{Extended version of the talk to be
published in the proceedings of the Mexican Meeting on Exact
Solutions and Scalar Fields in Gravity. M\'exico, 1-6 October, 2000. }
\footnotetext[2]{E-mail: cterrero@fis.cinvestav.mx}
\footnotetext[3]{Email: dschwarz@hep.itp.tuwien.ac.at}
\footnotetext[4]{Email: aagarcia@fis.cinvestav.mx}

\begin{abstract}
We present a new approximation scheme that allows us to increase the accuracy 
of analytical predictions of the power spectra of inflationary perturbations
for two specific classes of inflationary models. Among these models are
chaotic inflation with a monomial potential, power-law inflation and natural 
inflation (inflation at a maximum). After reviewing the 
established first order results we calculate the amplitudes and spectral 
indices for these classes of models at higher orders in the slow-roll 
parameters for scalar and tensor perturbations.  
\end{abstract}

\section{Introduction}
\label{sec:Intro}

Inflationary cosmology \cite{inflation} is facing exciting times 
due to a new generation of ground and satellite based experiments to 
be carried out (e.g., the SDSS, MAP and Planck experiments 
\cite{Exper}). Upcoming observations will allow us to determine 
the values of cosmological parameters with high confidence. To 
measure the values of these  parameters it is necessary to make assumptions 
on the initial conditions of the density fluctuations that evolved 
into the observed large scale structure and CMB anisotropies.

In the simplest inflationary scenario a single scalar field $\phi$ drives 
the accelerated expansion of the Universe. Cosmological perturbations are
generated by quantum fluctuations of this scalar field and of space-time.
Perturbations are adiabatic and gaussian and are characterized by their 
power spectra. Usually these spectra are described in terms of an amplitude at
a pivot scale and the spectral index at this scale. For general models, these 
quantities are difficult to compute exactly. The state-of-the-art 
in such calculations are the approximated expressions due to Stewart and Lyth 
\cite{StLy}, which are obtained up to next-to-leading order in terms of an 
expansion of the so-called slow-roll parameters. This expansion allows to 
approximate the solutions to the equations of motion by means of 
Bessel functions. A recent analysis of Wang et al. \cite{Wang} shows that
the Bessel function approximation cannot be further improved. We will show 
below that this result depends critically on the assumptions that are made on
the relative order of the various slow-roll parameters and that there are 
two regions in the slow-roll parameter space, where higher accuracy results 
may be obtained.

To reliably compare analytical predictions with measurements, 
an error in the theoretical calculations of some percents 
below the threshold confidence of observations is required. In 
Ref.~\cite{Lid4} it was shown that amplitudes of  
next-to-leading order power spectra can match the current level of 
observational precision. However, the error in the 
spectral index and the resulting net error in the multipole moments of 
the cosmic microwave background anisotropies might be large due to a long 
lever arm for wave numbers far away from the pivot scale \cite{MS}.
A clever choice of the pivot point is essential \cite{MS,Schwarz} for  
todays and future precision measurements. The slow-roll expressions 
as calculated in \cite{StLy} are not precise enough for the Planck experiment
\cite{MS}. Recently, Stewart and Gong \cite{GS} presented a new
method, also based on the slow-roll expansion, that allows to obtain 
analytical expressions at higher orders in the slow-roll parameters.

In this paper we show that the Bessel function approximation can be improved 
to higher orders, without being in conflict with the general argument of
\cite{Wang}. We 
calculate the amplitudes and indices of the scalar and tensorial perturbations
up to higher orders for two specific classes of inflationary models.
The first class contains all models that are `close' to power-law inflation,
one example is chaotic inflation with $V \propto \phi^\alpha$. Our second 
class of models is characterized by extremely slow rolling, an example is 
inflation near a maximum.

\section{The standard formulas}
\label{sec:SF}
 
Let us quickly review the derivation and results of the Bessel function 
approximation. The slow-roll parameters are defined as,
\cite{Lid3a},
\begin{equation}
\label{eq:SRP1} 
\epsilon(\phi) \equiv \frac{2}{\kappa}\left[\frac{H^\prime}{H}\right]^2,
\quad 
\eta(\phi) \equiv \frac{2}{\kappa}\frac{H^{\prime\prime}}{H},
\quad 
\xi(\phi) \equiv \frac{2}{\kappa}
\left(\frac{H^\prime H^{\prime\prime\prime}}{H^2}\right)^{1/2},
\end{equation}
with the equations of motion
\begin{equation}
\label{Cond1}
\frac{\dot{\epsilon}}{H}=2\epsilon(\epsilon-\eta)
\, ,
\quad
\frac{\dot{\eta}}{H}=\epsilon\eta-\xi^2
\, .
\end{equation}
$H$ is the Hubble rate, dot and prime denote derivatives with respect to 
cosmic time and $\phi$, $\kappa = 8\pi/m_{\rm Pl}^2$,  
and $m_{\rm Pl}$ is the Planck mass.
By definition $\epsilon\geq0$ and it
has to be less than unity for inflation to proceed. 

\subsection{Amplitudes of inflationary perturbations}
\label{eq:SS}

We call {\it{standard}} those formulas which
are considered to be the state-of-the-art in the analytical calculation
of perturbations spectra, i.e, those obtained by Stewart and Lyth 
\cite{StLy}.

The general expression for the spectrum of the curvature perturbations is
\cite{StLy}
\begin{equation}
{\mathcal P_R}^{1/2}(k)=\sqrt{\frac{k^3}{2\pi^2}}\left|\frac{u_k}{z}\right|
\, .  
\label{PrG}
\end{equation}
$u_k(\tau)$ are solutions of the mode equation \cite{StLy,Mukha}
\begin{equation}
\frac{d^2u_k}{d\tau^2}+\left(k^2-\frac{1}{z}\frac{d^2z}{d\tau^2}\right)u_k=0
\, ,  
\label{SchEq}
\end{equation}
where $\tau$ is the conformal time and $z$ is defined as 
$z\equiv a\dot{\phi}/H$.
The potential of the mode equation (\ref{SchEq}) reads \cite{StLy,Lid3a}
\begin{equation}
\frac{1}{z}\frac{d^2z}{d\tau^2}=2a^2H^2\left(1+\epsilon-
\frac{3}{2}\eta+\epsilon^2-2\epsilon\eta+\frac{1}{2}\eta^2+
\frac{1}{2}\xi^2\right)
\, .  
\label{eq:ztt}
\end{equation}
Despite its appearance as an expansion in slow-roll parameters, 
Eq.~(\ref{eq:ztt}) is an exact expression.

The crucial point in the Stewart and Lyth calculations is to use the 
solution for power-law inflation (where the slow-roll parameters are 
constant and equal each other) as a pivot expression to look for
a general solution in terms of a slow-roll expansion. An answer to whether 
the slow-roll parameters can be regarded as constants that differ from each 
other is found it looking at the exact equations of motion
(\ref{Cond1}). 

The standard procedure is to consider the slow-roll parameters so small 
that second order terms in any expression can be neglected. Now $aH$ in 
(\ref{eq:ztt}) can be replaced with help of 
$(aH)^{-1} \simeq - \tau (1 - \epsilon)$. 
Then, according with Eq.~(\ref{Cond1}) the 
slow-roll parameters can be fairly regarded as constants and Eq.~(\ref{SchEq})
becomes a Bessel equation readily solved. From that solution
the scalar amplitudes are written as \cite{StLy}
\begin{equation}
{\mathcal P_R}^{1/2}(k)=
2^{\nu-\frac{1}{2}}\frac{\Gamma(\nu)}{\Gamma(\frac{3}{2})}
(1-\epsilon)^{\nu-\frac{1}{2}}\frac{1}{m^2_{Pl}}
\left.\frac{H^2}{\left|H^\prime\right|}\right |_{k=aH}
\, ,
\label{PrPL}
\end{equation}
where $\nu$ is given by
\begin{equation}
\nu=\frac{1+\epsilon-\eta}{1-\epsilon}+\frac{1}{2}
\, ,
\label{nu}
\end{equation}
and $k$ is the wavenumber corresponding to the scale matching the Hubble
radius.
Expanding solution (\ref{PrPL}) with $\nu$ given by 
$\nu=3/2+2\epsilon-\eta$
and truncating the results to first order in $\epsilon$ and $\eta$ the
standard general expression for the scalar spectrum is obtained,
\begin{equation}
{\mathcal P_R}^{1/2}= 
\frac{\kappa}{4\pi}\left[1-(2C+1)\epsilon+C\eta\right]
\left.\frac{H^2}{\left|H^\prime\right|}\right |_{k=aH}
\, ,
\label{eq:PrSL}
\end{equation}
where $C=-2+\ln 2+\gamma\simeq -0.73$ is a numerical constant, and 
$\gamma$ is the Euler constant that arises when expanding the Gamma function.
Eq.~(\ref{eq:PrSL}) is called the next-to-leading order expression for the 
spectrum amplitudes of scalar perturbations, and from 
it the leading 
order is recovered by neglecting first order terms for $\epsilon$ and $\eta$.

The corresponding equation of motion for the tensorial modes is
\begin{equation}
\frac{d^2v_k}{d\tau^2}+\left(k^2-\frac{1}{a}\frac{d^2a}{d\tau^2}\right)v_k=0
\, ,  
\label{TSchEq}
\end{equation}
where,
\begin{equation}
\frac{1}{a}\frac{d^2a}{d\tau^2}=2a^2H^2\left(1-\frac12\epsilon
\right)
\, .  
\label{att}
\end{equation}
Neglecting any order of $\epsilon$ higher than the first one, 
Eq.~(\ref{att}) can be written as 
\begin{equation}
\frac{1}{a}\frac{d^2a}{d\tau^2}=\frac{1}{\tau^2}\left(\mu^2-\frac14\right)
\, ,
\label{att1}
\end{equation}
where,
\begin{equation}
\mu=\frac{1}{1-\epsilon}+\frac{1}{2} \simeq \frac32 + \epsilon
\, .
\label{mu}
\end{equation}
This way, Eq.~(\ref{TSchEq}) can be also approximated as a Bessel equation 
with solution,
\begin{equation}
{\mathcal P}_g^{1/2}(k)=2^{\nu-\frac{3}{2}}\frac{\Gamma(\nu)}
{\Gamma(\frac{3}{2})}
(1-\epsilon)^{\nu-\frac{1}{2}}
\left.\frac{\sqrt{2\kappa}}{\pi} H \right |_{k=aH}
\, .
\label{TPrPL}
\end{equation}
Substituting $\mu$ given by Eq.~(\ref{mu}) in Eq.~(\ref{TPrPL}), expanding on 
$\epsilon$ and truncating to first order, the Stewart-Lyth next-to-leading 
order result is obtained
\begin{equation}
{\mathcal P}_g^{1/2}= 
\left[1-(C+1)\epsilon\right]
\left. \frac{\sqrt{2\kappa}}{\pi} H \right|_{k=aH}
\, .
\label{eq:PgSL}
\end{equation}
The leading order equation is recovered by neglecting $\epsilon$.

\subsection{The spectral indices}  
\label{sec:SI}

To derive the expressions for the spectral indices we introduce here a 
variation of the standard procedure that has the advantage to allow 
a careful bookkeeping of the order of slow-roll expressions.

First, let us assume the following ansatz for the power spectrum of general
inflationary models,
\begin{equation}
{\mathcal P_R}^{1/2}=h(\epsilon, \eta, \xi)f(\epsilon, \eta, \xi) \, ,
\label{eq:anzats}
\end{equation}
where $f(\epsilon, \eta, \xi)$ may be written as a 
Taylor expansion
while $h(\epsilon, \eta, \xi)$ is a general function which
may be singular or discontinuous at $\epsilon=0$. Note that any 
function can be decomposed into this form.
We must also consider the following function of the slow-roll parameters:
\begin{equation}
\frac{\kappa}{2}\frac{H}{H^\prime}\frac{d\phi}{d\ln k} = - \frac{1}{1-\epsilon}
\equiv g(\epsilon, \eta, \xi)
\, .
\label{eq:g}
\end{equation}
In this expression all of the parameters are to be evaluated at values of 
$\phi$ corresponding to $k=aH$. The 
functions $f$ and $g$ can be expanded in Taylor series,
\begin{eqnarray}
\nonumber
f(\epsilon, \eta, \xi)&=& a_{00} + a_{10}\epsilon + a_{20}\eta 
+ a_{30}\xi \\
&+& a_{11}\epsilon^2 + a_{12}\epsilon\eta + a_{13}\epsilon\xi 
+ a_{22}\eta^2 + a_{23}\eta\xi + a_{33}\xi^2 
\cdots 
\, ,
\label{eq:S1} \\
g(\epsilon, \eta, \xi)&=& - (1 + \epsilon + \epsilon^2 + \cdots)
\, .
\label{eq:S2}
\end{eqnarray}
We proceed with the derivation of the equations for the scalar spectral
index
\begin{equation}
n_S(k)-1\equiv\frac{d\ln {\mathcal P_R}}{d\ln k}
\, .
\label{ns}
\end{equation}
{}From Eq.~(\ref{eq:anzats}) we obtain
\[
\frac{d\ln {\mathcal P_R}^{1/2}}{d\phi}= 
\frac{d\ln h(\epsilon, \eta, \xi) }{d\phi} +
\frac{d}{d\phi}\left[P(\epsilon, \eta, \xi) - 
\frac{P^2(\epsilon, \eta, \xi)}{2} + \frac{P^3(\epsilon, \eta, \xi)}{3} -
\cdots \right]
\, ,
\]  
where 
\[P(\epsilon, \eta, \xi)=  \tilde{a}_{10}\epsilon + \tilde{a}_{20}\eta 
+ \tilde{a}_{30}\xi + \tilde{a}_{11}\epsilon^2 + \tilde{a}_{12}\epsilon\eta 
+ \tilde{a}_{13}\epsilon\xi 
+ \tilde{a}_{22}\eta^2 + \tilde{a}_{23}\eta\xi + \tilde{a}_{33}\xi^2 
+ \cdots \, ,\]
and $\tilde{a}_{ij}\equiv a_{ij}/a_{00}$. After differentiation,
\begin{equation}
\frac{d\ln {\mathcal P_R}^{1/2}}{d\phi}= \frac{1}{h}\frac{dh}{d\phi} +
\left(1-P(\epsilon, \eta, \xi)+P^2(\epsilon, \eta, \xi) - \cdots \right)
P^\prime(\epsilon, \eta, \xi, \epsilon^\prime, \eta^\prime, \xi^\prime)
\, ,  
\label{eq:dlnAdphi3}
\end{equation}
where,
\[
P^\prime\equiv
\tilde{a}_{10}\epsilon^\prime + \tilde{a}_{20}\eta^\prime 
+ \tilde{a}_{30}\xi^\prime + 2\tilde{a}_{11}\epsilon\epsilon^\prime + 
\tilde{a}_{12}\left(\epsilon\eta\right)^\prime 
+ \tilde{a}_{13}\left(\epsilon\xi\right)^\prime 
+ 2\tilde{a}_{22}\eta\eta^\prime  + \tilde{a}_{23}\left(\eta\xi\right)^\prime 
 + 2\tilde{a}_{33}\xi\xi^\prime 
\]
plus higher order derivatives.
Using Eqs.~(\ref{eq:g}) and (\ref{eq:S2}),
and the definitions of the slow-roll parameters (\ref{eq:SRP1}) we obtain that
\begin{equation}
\frac{n_S-1}{2}= 
\left[ \sqrt{\frac{2}{\kappa}}\sqrt{\epsilon}\frac{h^\prime}{h} +
\left(1-P+P^2 - \cdots \right)P^\prime 
\right]
(1+\epsilon+\epsilon^2
+\epsilon^3 + \cdots)
\, ,  
\label{eq:dlnAdlnk2}
\end{equation}
where $P^\prime$ is written now as,
\begin{equation}
P^\prime = 2\tilde{a}_{10}\epsilon(\epsilon-\eta) + 
\tilde{a}_{20}(\epsilon\eta-\xi^2) + 
\tilde{a}_{30}\sqrt{\frac{2}{\kappa}\epsilon}\,\xi^\prime +
4\tilde{a}_{11}\epsilon^2(\epsilon-\eta) + \cdots
+ 2\tilde{a}_{22}\eta(\epsilon\eta-\xi^2) \cdots  
\, .  
\label{Pprime}
\end{equation}
Expression (\ref{eq:dlnAdlnk2}) can be used to any order whenever information 
on the function $h$ and the coefficients of the expansion (\ref{eq:S1}) is
available. The standard result of Stewart and Lyth, Eq.~(\ref{eq:PrSL}),
has been tested as a reliable approximation for the scalar spectrum of
general inflationary models \cite{Lid4}. Then, it is reasonable to assume
\begin{eqnarray}
\label{eq:hSL}
h(\epsilon, \eta, \xi) = \frac{\kappa}{4\pi}
\left.\frac{H^2}{\left|H^\prime\right|}\right |_{k=aH}&=& 
\frac{1}{2\pi}\sqrt{\frac{\kappa}{2}}\frac{H}{\sqrt{\epsilon}}
\, ,
\\
\label{a00}
a_{00}= 1
\, , \quad
a_{10}= -(2C+1)
\, , \quad
a_{20}&=& C \, .
\end{eqnarray}
With these assumptions, and ignoring the assumptions (constant slow-roll 
parameters) used to obtain Eq.~(\ref{eq:PrSL}), it is obtained that
\begin{equation}
\sqrt{\frac{2}{\kappa}}\sqrt{\epsilon}\frac{h^\prime}{h} =
-2\epsilon + \eta
\, ,
\label{hphs}
\end{equation}
and to second order (the order for which information on the coefficients for 
(\ref{Pprime}) is available from Eq.~(\ref{eq:PrSL})), expression 
(\ref{eq:dlnAdlnk2}) reduces to 
\begin{equation}
n_S-1= -4\epsilon + 2\eta - 8(C+1)\epsilon^2 + 2(5C+3)\epsilon\eta - 2C\xi^2
\, ,
\label{SLns}
\end{equation}
which is the standard result for the scalar spectral index.

The equation of the tensorial index can be derived in a similar way.
By definition 
\begin{equation}
n_T(k)\equiv\frac{d\ln {\mathcal P}_g}{d\ln k}
\, .
\label{nT}
\end{equation}
The standard equation for tensorial modes is given by
Eq.~({\ref{eq:PgSL}). If we now use expression (\ref{eq:dlnAdlnk2}) 
substituting $n_S-1$ by $n_T$ and 
\begin{eqnarray}
\label{eq:hTSL}
h(\epsilon, \eta, \xi) &=& \frac{\sqrt{2\kappa}}{\pi} H
\, ,
\\
\label{aT00}
a_{00}= 1
\, , \quad
a_{10}&=& -(C+1)
\, ,
\end{eqnarray}
then we obtain
\begin{equation}
\sqrt{\frac{2}{\kappa}}\sqrt{\epsilon}\frac{h^\prime}{h} =
-\epsilon 
\, ,
\label{hphT}
\end{equation}
and the standard expression,
\begin{equation}
n_T= -2\epsilon - 2(2C+3)\epsilon^2 + 4(C+1)\epsilon\eta
\, .
\label{SLnT}
\end{equation}

\section{Generalizing the Bessel approximation}
\label{sec:Gen}

As can be observed from Eq.~(\ref{Cond1}), the standard
assumption used to approximate the slow-roll parameters as constants can 
not be used beyond the linear term in the slow-roll expansion. Hence, from
this point of view, the feasibility of using the Bessel equation to calculate
the power spectra is limited to this order. Nevertheless, what it is actually
needed is the right hand sides of Eqs.~(\ref{Cond1}) 
being negligible. That can also be achieved if
$\left|\upsilon\right| \equiv \left|\epsilon-\eta\right| \ll \epsilon$ (we 
shall call this the power-law approximation) or if 
$\epsilon\ll\left|\upsilon\right|$ (we call this the extreme slow-roll 
approximation).
In both cases we require 
$\left|\dot{\upsilon} H^{-1}\right| = \left|\xi^2 - \epsilon^2 + 
3 \epsilon \upsilon\right| \ll {\rm min}(\epsilon, \left|\upsilon\right|)$.
No discrepancy arises with the conclusions in Ref.~\cite{Wang}, where
\[
\frac{d\ln(\epsilon)}{dN} = -2\upsilon \sim \epsilon 
\]
is assumed, $N$ being the number of e-foldings. As explained above, this 
condition is not fulfilled in neither of our approximations.

\subsection{Power-law approximation}

Power-law inflation is the model which gives rise to the commonly used 
power-law shape of the primordial spectra, although
not always properly implemented \cite{Schwarz}. The assumption of a 
power-law shape of the spectrum has been successful in describing
large scale structure from the scales probed by the cosmic microwave 
background to the scales probed by redshift surveys. It is reasonable to
expect that the actual model behind the inflationary perturbations has 
a strong similarity with power-law inflation. For this class of potentials 
the precision of the power spectra calculation can be increased while still 
using the Bessel approximation. If we have a model with 
$\upsilon\propto\epsilon^n$, then, according with Eq.~(\ref{Cond1}), 
the first slow-roll parameter can be
considered as constant if terms like $\epsilon^{n+1}$ and with higher orders 
are neglected. 
These considerations, plus the condition $\epsilon^2\sim\xi^2$, imply the 
right hand side
of the second equation in (\ref{Cond1}) to be also negligible and, this way, $\eta$ can be
regarded as a constant too. The higher the order in the slow-roll parameters, 
the smaller should
be the difference between them, finally leading, for infinite order in the 
parameters, to the case of power-law inflation. That means that this approach 
to the problem of calculating the spectra for
 more general inflationary models is in fact an expansion around the 
power-law solution. An example is given by chaotic inflation with the 
potential $V \propto \phi^\alpha$ with $\alpha > 2$. In this case the 
slow-roll 
parameters are given by $\epsilon \simeq \alpha/(4 N)$ and 
$\upsilon \simeq 1/(2 N)$ \cite{inflation}, i.e., $\epsilon > \upsilon$. 
$N$ denotes the number of e-folds of inflation.

Let us proceed with the calculations. In fact all that has to
be done is to repeat the calculations of the previous sections keeping the 
desired order in the slow-roll expansions and neglecting all 
the terms similar to the right hand sides of Eqs.~(\ref{Cond1}).
For example, the next-to-next-to-leading order expression for the scalar
power spectrum will be obtained from solution (\ref{PrPL}) but now with
$\nu$ given by $\nu= 3/2 +\epsilon+\upsilon+\epsilon^2$.
Expanding and keeping terms up to second order, the final expression is
\begin{equation}
{\mathcal P_R}^{1/2}(k)= \frac{\kappa}{4\pi}
\left[1-(C+1)\epsilon-C\upsilon+
(B-1)\epsilon^2\right]
\left.\frac{H^2}{\left|H^\prime\right|}\right |_{k=aH}
\, ,
\label{PrGST}
\end{equation}
where $B=-2+C^2/2+\pi^2/4\simeq 0.73$.  
This result is consistent with the result obtained in Ref.\ \cite{GS}. 
Eq.~(\ref{eq:PrSL}) can be readily 
recovered from Eq.~(\ref{PrGST}) by neglecting second order terms. 
For the corresponding expression of the tensorial power spectrum, $\mu$ must
be $\mu=3/2+\epsilon+\epsilon^2$
and,
\begin{equation}
{\mathcal P}_g^{1/2}= 
\left[1-(C+1)\epsilon + (B-1)\epsilon^2\right]
\left. \frac{\sqrt{2\kappa}}{\pi} H \right|_{k=aH}
\, .
\label{eq:PgSLnnlo}
\end{equation}
Equation (\ref{eq:PgSL}) is obtained from Eq.~(\ref{eq:PgSLnnlo}) by
neglecting second order terms of $\epsilon$.

\subsection{Extreme slow-roll approximation}

There is no reason to reject the possibility of an inflationary
model with $|\upsilon| \gg\epsilon$. As we shall see later,
some important models belong to this class. For these cases, the accuracy of 
the calculations
of the amplitudes can also be increased. We shall focus on the 
next-to-next-to-leading order. With regards of 
conditions (\ref{Cond1}), we neglect terms like 
$\epsilon^2$, $\epsilon\upsilon$ and $\xi^2 - \epsilon^2$ but we 
keep $\upsilon^2$ terms.
Repeating the calculations under this set of assumptions we obtain for the
amplitudes of the scalar spectrum:
\begin{equation}
{\mathcal P_R}^{1/2}= 
\frac{\kappa}{4\pi}\left[1-(C+1)\epsilon-C\upsilon 
+ B\upsilon^2\right]
\left.\frac{H^2}{\left|H^\prime\right|}\right |_{k=aH}
\, .
\label{eq:PrGSL}
\end{equation} 
which again agrees with the result from \cite{GS} in the appropriate
limit.
For tensorial perturbations we note that the Bessel function index
(\ref{mu}) does not depend on $\upsilon$ hence, no term like 
$\upsilon^2$ will arise
at any moment of the calculations. This way, 
the expression for the tensorial amplitudes is 
given by Eq.~(\ref{eq:PgSL}).

\subsection{The spectral indices}

In general, if the approximation neglecting the right hand sides of
Eqs.~(\ref{Cond1}) is consistently taken into account, expressions for 
the spectral indices at any order $n$ are easy
to be derived noting that in these cases expression 
(\ref{Pprime}) always vanishes and, 
\begin{equation}
\frac{n_i}{2}= 
\left[ \sqrt{\frac{2}{\kappa}}\sqrt{\epsilon}\frac{h_i^\prime}{h_i}
\right]
(1+\epsilon+\epsilon^2 + \dots + \epsilon^n)
\, ,  
\label{eq:dlnAdlnk2nnlo}
\end{equation}   
where, $n_i$ is $n_S-1$ or $n_T$ and $h_i$ is correspondingly given by
Eqs.(\ref{eq:hSL}) and (\ref{eq:hTSL}).
This way it is obtained 
\begin{eqnarray}
\label{nsnnlo}
n_S(k)-1&=&(-4\epsilon+2\eta)(1+\epsilon+\epsilon^2
+ \dots + \epsilon^n) \nonumber \\
&=& -2\upsilon - 2\epsilon(1+\epsilon + \epsilon^2 + \dots + \epsilon^n)
\, , \\
\label{nTnnlo}
n_T(k)&=&-2\epsilon(1+\epsilon+\epsilon^2+\dots+\epsilon^n)
\, .
\end{eqnarray}

\subsubsection{Relaxing the assumptions}

In the same way as the spectral indices (\ref{SLns}) and (\ref{SLnT}) 
have been derived up to second order in the slow-roll parameters, we can now 
relax our assumptions on the slow-roll parameters and include higher orders
beyond those already included in (\ref{nsnnlo}) and (\ref{nTnnlo}).

Let us first consider the case of the power-law approximation.
The amplitudes are given by Eqs.~(\ref{PrGST}) and (\ref{eq:PgSLnnlo}). We 
now drop the assumption that $\epsilon, \upsilon$, and $\epsilon^2$ are 
constant and keep all the contributions from the derivatives of these terms, 
i.e., $\epsilon \upsilon, \dot{\upsilon}/H, \epsilon^2 \upsilon$. However,
we have to make sure that no derivatives of terms that have been neglected 
in the amplitude show up, thus we have to neglect the derivatives of 
$\epsilon \upsilon, \upsilon^2$, and $\dot{\upsilon}/H$, which gives rise to 
the conditions $\epsilon \upsilon^2 = - \epsilon \dot{\upsilon}/(2H), 
\upsilon \dot{\upsilon}/H = 0$, and 
$[{\rm d}(\dot{\upsilon}/H)/{\rm d}t]/H = 0$. This means that we are allowed 
to keep the terms $\epsilon^3, \epsilon^2  \upsilon$, and 
$\epsilon\dot{\upsilon}/H$ beyond the standard expression for the spectral 
indices.
 
Consistently with our approach, we must write the expressions for $P$ 
and $P^{\prime}$ in terms of our basic parameters $\epsilon$ and $\upsilon$. 
These expressions are,
\begin{eqnarray}
\label{Pour}
P &=& \tilde{a}_{10}\epsilon + \tilde{a}_{20}\upsilon 
+ \tilde{a}_{11}\epsilon^2 + \tilde{a}_{12}\epsilon\upsilon 
+ \tilde{a}_{22}\upsilon^2 + \cdots \, ,\\
\label{Pprimeour}
P^\prime &=& 2\tilde{a}_{10}\epsilon\upsilon + 
\tilde{a}_{20}\dot{\upsilon}/H 
\nonumber \\
& & +\, 4\tilde{a}_{11}\epsilon^2\upsilon 
+ \tilde{a}_{12}\epsilon(2\upsilon^2 + \dot{\upsilon}/H)
+ 2\tilde{a}_{22}\upsilon \dot{\upsilon}/H + \cdots  
\, ,  
\end{eqnarray}  
where $\sqrt{2\epsilon/\kappa}\upsilon^\prime = \dot{\upsilon}/H 
= 2\epsilon^2 - 3\epsilon\eta + \xi^2$.
For the scalar contribution in the power-law approximation 
\begin{equation}
\label{a00nnlo}
\tilde{a}_{10}= -(C+1)
\, ,
\quad
\tilde{a}_{20}= -C
\, ,
\quad
\tilde{a}_{11}= B-1
\, .
\end{equation}
Using Eqs.~(\ref{Pour}) and (\ref{Pprimeour}), and taking into account the 
relaxed approximations discussed above, expression (\ref{eq:dlnAdlnk2}) 
reduces to 
\begin{eqnarray}
n_S-1&=& -2\epsilon - 2\upsilon - 2\epsilon^2 - 2(2C+3)\epsilon\upsilon 
- 2C\dot{\upsilon}/H 
\nonumber \\
& & 
- 2\epsilon^3 - 2(6C + 17 - \pi^2)\epsilon^2\upsilon 
- 2C\epsilon\dot{\upsilon}/H \, .
\label{GPLnsour}
\end{eqnarray}
Correspondingly, for the tensorial index
\begin{equation}
\label{a00nnlonT}
\tilde{a}_{10}= -( C+1)
\, ,
\quad
\tilde{a}_{11}= B-1
\, ,
\end{equation}
and
\begin{eqnarray}
n_T &=& -2\epsilon - 2\epsilon^2 - 4(C+1)\epsilon\upsilon
- 2\epsilon^3 - 2(6C + 16 - \pi^2)\epsilon^2\upsilon \, .
\label{SLnTnnlo1}
\end{eqnarray}

In a similar manner we can derive spectral indices for the extreme
slow-roll approximation. Now we have to make use of the conditions 
$\epsilon^2 \upsilon = 0, \epsilon \dot{\upsilon}/H = -2 \epsilon \upsilon^2$,
and $[{\rm d}(\dot{\upsilon}/H)/{\rm d}t]/H = 0$. We have now for the scalars 
\begin{equation}
\label{a00nnloGSL}
\tilde{a}_{10}= -(C+1)
\, ,
\quad
\tilde{a}_{20}= -C
\, ,
\quad
\tilde{a}_{22}= B
\, ,
\end{equation}
and the spectral index reads, 
\begin{eqnarray}
n_S-1 &=& -2\epsilon - 2\upsilon - 2\epsilon^2 - 2(2C+3)\epsilon\upsilon 
- 2C \dot{\upsilon}/H 
\nonumber \\
&& - 2\epsilon^3 
+ 4 C \epsilon \upsilon^2 - (8 -\pi^2) \upsilon \dot{\upsilon}/H
\, ,
\label{GSRnsour}
\end{eqnarray}
while the tensorial index is identical to the standard second order slow-roll 
result,
\begin{eqnarray}
n_T &=& -2\epsilon - 2\epsilon^2 - 4(C+1)\epsilon\upsilon \,  .
\end{eqnarray}

To conclude this section let us not that our expressions for the spectral 
indices are in full agreement with the results of Stewart and Gong \cite{GS}, 
taking the corresponding approximations into account. This is a quite 
nontrivial test of our results and of the results of Ref. \cite{GS}.
 
\section{Testing the expressions}
\label{sec:Test}

To test all the expressions presented in this paper, we will use the well 
known exact results for amplitudes and indices calculations in the cases of
power-law and natural inflation.

Power-law inflation \cite{Lucchin} is an inflationary scenario where,
\begin{equation}
\label{eq:PL}
a(t) \propto t^p \, ,
\quad 
H(\phi) \propto \exp\left(-\sqrt{\frac{\kappa}{2p}}\,\phi\right),
\quad
V(\phi) \propto \exp\left(-\sqrt{\frac{2\kappa}{p}}\,\phi\right),
\end{equation}
with $p$ being a positive constant. It follows from (\ref{eq:SRP1})
that in this case the slow-roll
parameters are constant and equal each other, 
\begin{equation}
\epsilon=\eta=\xi=1/p \, .
\label{PLcond}
\end{equation} 
Thus, this is the limit case of the power-law approximation. Note
that condition $\epsilon<1$ implies $p>1$.

For this model, the power spectrum of scalar perturbations is given by
\cite{LySt}
\begin{equation}
{\mathcal{P}_R}^{1/2}(k)=\frac{2^{\nu-\frac{1}{2}}}{m^2_{Pl}}
\frac{\Gamma(\nu)}{\Gamma(\frac{3}{2})}
\left(1-\frac{1}{p}\right)^{\nu-\frac{1}{2}}
\left.\frac{H^2}{\left|H^\prime\right|}\right |_{k=aH} \, ,
\label{eq:PrPL}
\end{equation}
with
\[
\nu\equiv\frac{3}{2}+\frac{1}{p-1}.
\]
Expanding up to second order for $p \gg 1$ it is obtained that
\begin{equation}
{\mathcal P_R}^{1/2}(k)= \frac{\kappa}{4\pi}
\left[1-(C+1)\frac{1}{p} +
(B-1)\frac{1}{p^2}\right]
\left.\frac{H^2}{\left|H^\prime\right|}\right |_{k=aH}
\, ,
\label{GPLTest}
\end{equation}
in full correspondence with Eq.~(\ref{PrGST}) when relations (\ref{PLcond}) are
taken into account.
Testing the tensorial amplitudes can be easily done by checking that 
Eqs.~(\ref{PrGST}) and (\ref{eq:PgSLnnlo}) indeed satisfy the relation between
amplitudes characteristic of power-law inflation, i.e.,
\begin{equation}
{{\mathcal P}_g}^{1/2}(k)=\frac{4}{\sqrt{p}}{\mathcal P_R}^{1/2}(k)
\, .
\label{eq:PgPL}
\end{equation}
For the spectral indices one can see that substitution of
relations (\ref{PLcond}) in Eqs.~(\ref{nsnnlo}) and (\ref{nTnnlo}), as well as 
in Eqs.~(\ref{GPLnsour}) and (\ref{SLnTnnlo1}), yield
\begin{equation}
\frac{n_S-1}{2}= \frac{n_T}{2} =
-\frac{1}{p}\left(1+\frac{1}{p}+\frac{1}{p^2}
+\cdots\right)=\frac{1}{1-p}
\, ,
\label{eq:PLns}
\end{equation}
the corresponding relation between indices in power-law inflation.

Another inflationary scenario where precise predictions for the spectra 
amplitudes and indices can be done is natural inflation \cite{Freese}. In
this case, the potential is given by,
\begin{equation}
V = \Lambda^4 \left[1\pm\cos(\frac{\phi}{f})\right] \, ,
\label{NIV}
\end{equation}
where $\Lambda$ and $f$ are mass scales. 
For simplicity we choose the plus sign in the remaining calculations.
The point here is to analyze inflation near the origin so that the small-angle
approximation applies, i.e., $\phi\ll f$. In this case the slow-roll 
parameters are,
\begin{eqnarray}
\label{NIep}
\epsilon &=& \frac34 \kappa \left(\sqrt{1+\frac23 \frac{1}{\kappa f^2}}-1\right)
 \phi^2 \simeq 0
 \, , \\
\eta &=& -\frac32 \left(\sqrt{1+\frac23 \frac{1}{\kappa f^2}}-1\right) \, ,\\
\label{NIeta}
\xi^2 &\simeq& 0 \, ,\\
\label{NIxi}
\upsilon &\simeq& -\eta 
\, .
\label{NIup}
\end{eqnarray}
As it can be observed, in this approximation natural inflation belongs to the
class of models that fulfill the conditions of the extreme slow-roll 
approximation.
With the values of the parameters given above, Eq.~(\ref{PrPL}) can be used 
with the corresponding $\nu = 3/2 + \upsilon$.
After expanding and truncating at the proper order
\begin{equation}
{\mathcal P_R}^{1/2}= 
\frac{\kappa}{4\pi}\left[1-C\upsilon 
+ B\upsilon^2\right]
\left.\frac{H^2}{\left|H^\prime\right|}\right |_{k=aH}
\, 
\label{eq:GSLtest}
\end{equation}
is obtained, consistent with Eq.~(\ref{eq:PrGSL}). With this same set of 
assumptions, Eq.~(\ref{nsnnlo}) and (\ref{GSRnsour}) are reduced to
\begin{equation}
n_S - 1 = 2\eta
\, ,
\label{NIns}
\end{equation}
the well known result for natural inflation.

\section{Conclusions}
\label{sec:Concl}

In this paper we introduced a simple procedure which allows to improve 
the precision of the predictions of inflationary perturbations 
for two specific classes of inflationary models. 
The method is based on neglecting the evolution in time of the slow-roll 
parameters $\epsilon$ and $\upsilon = \epsilon - \eta$. All monomials of 
the slow-roll parameters $\epsilon$ and $\upsilon$ that are larger than 
${\rm max}(\dot{\epsilon}/H,\dot{\upsilon}/H)$ can be taken into account,
the rest is dropped consistently.  Two cases arise. 
In the first one, $\epsilon$ is greater than $\upsilon$. We called this 
approach power-law approximation.
In the opposite case, the value of $\epsilon$ is smaller  
than $\upsilon$ (extreme slow-roll approximation).
These approximations plus the standard 
approximation cover a large space of inflationary models.

We briefly discussed the standard approach to the derivation of the spectral 
indices. It was noted that the information contained in the expressions for 
the power spectra amplitudes obtained within the Bessel formalism is used as
an approximation to derive the spectral indices for more general models. 
The obtained expressions have been tested against the exact 
results for power-law and natural inflation. 

\section*{Acknowledgement}

We want to thank Andrew Liddle for helpful discussions. D.J.S. thanks 
J\'er\^ome Martin for collaborations on closely related issues.
The work of C.A.T-E and A.A.G. is supported in part by the CONACyT 
grant 32138--E and the Sistema Nacional de Investigadores (SNI). 
D.J.S. would like to thank the Austrian Academy of Sciences for financial 
support.

\end{document}